\documentclass[letterpaper,onecolumn,accepted=2022-12-30]{quantumarticle}
\pdfoutput=1

\usepackage{bm}
\usepackage{amsfonts}
\usepackage{amsmath}
\usepackage{amssymb}
\usepackage{xcolor}
\usepackage{graphicx}
\usepackage{amsthm}
\usepackage{qtree}
\usepackage{tree-dvips}
\usepackage{float}
\usepackage{hyperref}
\usepackage[nameinlink]{cleveref}
\usepackage{braket}
\usepackage{mathrsfs}
\usepackage{tikz}
\usepackage{qcircuit}
\usepackage{dsfont}

\usepackage[numbers,sort&compress]{natbib}

\hypersetup{colorlinks={true},linkcolor={blue},citecolor=magenta}

\theoremstyle{plain}

\newcommand{\ketbra}[2]{\ket{#1}\!\bra{#2}}
\renewcommand{\cal}[1]{\mathcal{#1}}

\newcommand{\C}{\mathbb{C}}

\newcommand{\Tr}{\mathrm{Tr}}

\date{}
\title{An Improved Sample Complexity Lower Bound for (Fidelity) Quantum State Tomography}

\author{Henry Yuen}
\affiliation{Columbia University}

\begin{document}
\maketitle

\begin{abstract}
    We show that $\Omega(rd/\epsilon)$ copies of an unknown rank-$r$, dimension-$d$ quantum mixed state are necessary in order to learn a classical description with $1 - \epsilon$ fidelity. This improves upon the tomography lower bounds obtained by Haah et al. and Wright (when closeness is measured with respect to the fidelity function).
\end{abstract}

\section{Background}

We consider quantum state tomography: given $n$ copies of a mixed state $\rho$, output a classical description of a state $\sigma$ that is close to $\rho$. In this note we measure closeness between $\rho$ and $\sigma$ via their \emph{fidelity} $F(\rho,\sigma)$, defined as the supremum of $|\langle \varphi \mid \psi \rangle|^2$ over all purifications $\ket{\psi},\ket{\varphi}$ of $\rho,\sigma$ respectively\footnote{We note that there are two versions of fidelity considered in the literature; this is the \emph{squared} one.}. In \emph{fidelity tomography}, the goal is for the output state $\sigma$ to satisfy $F(\rho,\sigma) \geq 1 - \epsilon$.
Haah et al.~\cite{haah2017sample} showed that $n = O(rd \log(d/\epsilon) /\epsilon)$ is sufficient for fidelity tomography where $r$ is the rank of the density matrix $\rho$. They also proved a $n = \Omega \Big( \frac{rd}{ \delta^2 \log (d/r\delta)} \Big)$ lower bound for \emph{trace distance tomography}, where the goal is to output a state $\sigma$ whose trace distance $\| \rho - \sigma \|_1 = \frac{1}{2} \Tr ( | \rho - \sigma|)$ with $\rho$ is at most $\delta$. The Fuchs-van de Graff inequalities then imply a $n = \Omega \Big( \frac{rd}{ \epsilon \log (d/r\epsilon)} \Big)$ lower bound for fidelity tomography; thus their upper and lower bounds are tight up to logarithmic factors. %
O'Donnell and Wright~\cite{o2016efficient} proved that $n = O(rd/\delta^2)$ input samples are sufficient for trace distance tomography. 
In his PhD thesis, Wright~\cite{wright2016} showed that $n = \Omega(rd)$ samples are necessary for both fidelity and trace distance tomography; this is optimal when the desired trace distance error $\delta$ or infidelity $\epsilon$ is fixed to a constant, but otherwise is loose when the error is treated as a quantity going to zero. 



In this note we improve the lower bound of Haah et al.~\cite{haah2017sample} and Wright~\cite{wright2016} in the fidelity tomography case and show that $n = \Omega(rd/\epsilon)$ input samples are needed. 
It is natural to conjecture that the optimal bound for fidelity tomography is $n = \Theta(rd/\epsilon)$; however we leave obtaining a matching upper bound for future work.

\section{The argument}

We prove our lower bound via reduction to the \emph{pure state} tomography scenario, in which the input samples $\rho$ are guaranteed to be pure states (in other words, the rank of the density matrix is $1$). It was proved by Bru{\ss} and Macchiavello~\cite{bruss1999optimal} that $n=\Theta(d/\epsilon)$ samples are necessary and sufficient to achieve fidelity $1 - \epsilon$; this was based on a tight connection between optimal pure state tomography and optimal pure state cloning~\cite{werner1998optimal,keyl1999optimal}.

Suppose there is an algorithm $\cal{A}$ that, on input $n$ copies of a rank-$r$, dimension-$d$ mixed state $\rho$, outputs with high probability a classical description of a state $\sigma$ that has fidelity $1 - \epsilon$ with $\rho$. Then we use this algorithm to construct another algorithm $\cal{B}$ that performs tomography on \emph{pure}, dimension-$rd$ states using $O \Big (n + \frac{r^2}{\epsilon} \Big )$ input samples and achieves $1 - O(\epsilon)$ fidelity. The performance of algorithm $\cal{B}$ is subject to the bounds of Bru{\ss} and Macchiavello~\cite{bruss1999optimal} -- in other words, $\cal{B}$ must use at least $\Omega(rd/\epsilon)$ input samples. Thus it must be that
\[
    n = \Omega(rd/\epsilon) - O(r^2/\epsilon) = \Omega(rd/\epsilon)~,
\]
as desired. 

The algorithm $\cal{B}$ works as follows. Let $\ket{\psi}_{XY}$ denote the $rd$-dimensional pure input sample where $X$ denotes an $r$-dimensional register and $Y$ denotes a $d$-dimensional register. 
\begin{enumerate}
    \item The algorithm $\cal{B}$ takes $n$ input samples $\ket{\psi}_{XY}$ and traces out the $X$ register in each copy to obtain $n$ copies of a mixed state $\rho\in \C^{d \times d}$. 
    \item Run algorithm $\cal{A}$ on the $n$ copies of $\rho$ to obtain (with high probability) a classical description of a rank-$r$, dimension-$d$ state $\sigma$ that has fidelity $1 - \epsilon$ with $\rho$.
    \item Compute a classical description of the rank-$r$ projector $\Pi$ onto the support of $\sigma$.
    \item Take $O(r^2/\epsilon)$ additional copies of the input state $\ket{\psi}_{XY}$ and measure the $Y$ register of each copy using the projective measurement $\{\Pi, I - \Pi\}$, and keep the post-measurement states $\ket{\widetilde{\psi}}$ of the copies where the $\Pi$ outcome was obtained. 
    
    \item Use the tomography procedure of~\cite{bruss1999optimal} for dimension-$r^2$ pure states on the copies of $\ket{\widetilde{\psi}}$ where we treat the states as residing in the dimension-$r^2$ subspace
    \[
        \C^r \otimes \mathrm{supp}(\Pi) \subseteq \C^r \otimes \C^d~.
    \]
    Let $\ket{\varphi} \in \C^r \otimes \mathrm{supp}(\Pi)$ denote the result of this pure state tomography procedure. The algorithm $\cal{B}$ then outputs the classical description of $\ket{\varphi}$ as its estimation of $\ket{\psi}$.
\end{enumerate}

We analyze the algorithm $\cal{B}$. Let
\[
    \ket{\psi}_{XY} = \sum_{i=1}^r \lambda_i \ket{u_i} \otimes \ket{v_i}
\]
denote the Schmidt decomposition of $\ket{\psi}$ where $\{ \ket{u_1},\ldots,\ket{u_r} \}$ is an orthonormal basis for $\C^r$ and $\{ \ket{v_1},\ldots,\ket{v_r} \}$ is an orthogonal set of vectors in $\C^d$. We can then write $\rho$ as
\[
    \rho = \Tr_X(\ketbra{\psi}{\psi}) = \sum_{i=1}^r \lambda_i^2 \ketbra{v_i}{v_i}~.
\]
Note that $\rho$ is a rank-$r$ density matrix.

Let $\sigma = \sum_{i=1}^r \mu_i \ketbra{w_i}{w_i}$ denote the estimate produced by Step 2. By the guarantees of algorithm $\cal{A}$, we have that (with high probability) $F(\rho,\sigma) \geq 1 - \epsilon$. Let $\Pi = \sum_{i=1}^r \ketbra{w_i}{w_i}$ denote the projector onto the support of $\sigma$. By the definition of fidelity, there exists a purification $\ket{\varphi} \in \C^r \otimes \C^d$ of $\sigma$ such that 
\[
    F(\rho,\sigma) = | \langle \psi \mid \varphi \rangle |^2 \geq 1 - \epsilon~.
\]
On the other hand,
\begin{align*}
    | \langle \psi \mid \varphi \rangle |^2 = | \langle \psi \mid (I \otimes \Pi) \, \rvert \varphi \rangle |^2 \leq \langle \psi \lvert I \otimes \Pi \rvert \psi \rangle \cdot \langle \varphi \mid \varphi \rangle = \langle \psi \lvert I \otimes \Pi \rvert \psi \rangle
\end{align*}
where the inequality uses Cauchy-Schwarz. Let $\ket{\widetilde{\psi}} = (I \otimes \Pi) \ket{\psi} / \| I \otimes \Pi \ket{\psi} \|$, and observe that
\begin{align*}
    | \langle \widetilde{\psi} \mid \psi \rangle |^2
    &= \frac{1}{\| I \otimes \Pi \ket{\psi} \|^2} \, | \bra{\psi} I \otimes \Pi \ket{\psi} |^2 \\
    &= \frac{1}{ |\bra{\psi} I \otimes \Pi \ket{\psi}|}  \,\, | \bra{\psi} I \otimes \Pi \ket{\psi} |^2 \\
    &= | \bra{\psi} I \otimes \Pi \ket{\psi} | \geq 1 - \epsilon~.
\end{align*}
The number of copies of $\ket{\widetilde{\psi}}$ available in Step 5 is, with high probability, at least $\Omega(r^2/\epsilon)$. Thus the estimate $\ket{\varphi}$ computed by Step 5 will satisfy
\[
    | \langle \varphi \mid \widetilde{\psi} \rangle|^2 \geq 1 - \epsilon
\]
with high probability. 

We now turn to a simple geometric proposition: if $| \langle \widetilde{\psi} \mid \psi \rangle | \geq 1 - \eta$ and $| \langle \varphi \mid \widetilde{\psi} \rangle| \geq 1 - \eta$, then $| \langle \varphi \mid \psi \rangle| \geq 1 - 4\eta$. This is because $| \langle \widetilde{\psi} \mid \psi \rangle | \geq 1 - \eta$ implies
\[
    \| \ket{\widetilde{\psi}} - e^{i \alpha} \ket{\psi} \|^2 = 2 - 2 | \langle \widetilde{\psi} \mid \psi \rangle | \leq 2\eta
\]
where $e^{i\alpha}$ is a complex phase satisfying $| \langle \widetilde{\psi} \mid \psi \rangle | = e^{i \alpha}\langle \widetilde{\psi} \mid \psi \rangle $. 
Similarly $\| \ket{\widetilde{\psi}} - e^{i \beta} \ket{\varphi} \|^2 \leq 2\eta$ for some complex phase $e^{i \beta}$. Then by triangle inequality we have
\begin{align*}
    2 - 2|\langle \varphi \mid \psi \rangle| &\leq \| e^{i \beta} \ket{\varphi} - e^{i \alpha} \ket{\psi} \|^2 \\
    &\leq 2\| \ket{\widetilde{\psi}} - e^{i \alpha} \ket{\psi} \|^2 + 2 \| \ket{\widetilde{\psi}} - e^{i \beta} \ket{\varphi} \|^2 \leq 8 \eta~.
\end{align*}
The proposition follows via rearrangement.

Therefore with high probability, the estimate $\ket{\varphi}$ produced by algorithm $\cal{B}$ satisfies
\[
    | \langle \varphi \mid \psi \rangle|^2 \geq (1 - 8\epsilon)^2 \geq 1 - 16\epsilon~.
\]


\section{Conclusion}

We proved a $\Omega(rd/\epsilon)$ sample complexity lower bound for fidelity tomography for rank-$r$, dimension-$d$ mixed states where $1-\epsilon$ is the fidelity of the resulting estimate. This is proved via reduction to the $\Omega(d/\epsilon)$ lower bound for pure state tomography established by~\cite{bruss1999optimal}. In contrast, the lower bounds of~\cite{haah2017sample} and~\cite{wright2016} are based on communication complexity arguments. Natural questions include: (a) whether the upper bound for fidelity tomography can be improved to $O(rd/\epsilon)$, and (b) whether a $\Omega(rd/\delta^2)$ lower bound can be established for trace distance tomography. One obstacle to extending our argument to the trace distance setting is that we do not know whether applying the projection $\Pi$ to the state $\ket{\psi}$ (if $\Pi$ is the projector onto the support of a state $\sigma$ that is $\delta$-close to $\rho$ in trace distance) yields a state that is $O(\delta)$-close to $\ket{\psi}$ in trace distance. The Gentle Measurement Lemma~\cite{winter1999coding} implies that the post-measurement state is $O(\sqrt{\delta})$-close to $\ket{\psi}$; this ultimately yields a $\Omega(rd/\delta)$ lower bound on trace distance tomography, which we believe is not optimal.

\subsection*{Acknowledgments} We thank John Wright, Jeongwan Haah, and an anonymous reviewer for helpful feedback. This work was supported by AFOSR award FA9550-21-1-0040, NSF CAREER award CCF-2144219, and the Sloan Foundation.

\bibliographystyle{plainnat}
\bibliography{refs}

\end{document}